\documentstyle[epsf,12pt]{article}

\begin{document}

\title{ The short-time behavior of kinetic spherical model
with long-ranged interactions
\thanks{Supported by the National Natural Science
Foundation of China under the project 19772074 and by the Deutsche Forschungsgemeinschaft under the
project Schu 95/9-1.}
}
\author{\normalsize \sf Yuan Chen $^{a}$,Shuohong Guo$^{a}$,
Zhibing Li$^{a,b}$ and Aijun Ye$^{a}$   \\
\normalsize $^{a}$Department of Physics,  Zhongshan University,
Guangzhou 510275, China \\
\normalsize $^{b}$ Associate Member of the ICTP, Italy} 
\date{}

\maketitle
Shortened version of the title: The short-time behavior of kinetic 
spherical model

\vskip 1.5pc
\begin{abstract}
The kinetic spherical model with long-ranged
interactions and an arbitrary initial order $m_{0}$
quenched from a very high temperature to $T$ $(\leq T_{c})$ is solved. 
In the short-time regime, the bulk order increases with a 
power law in both
the critical and phase-ordering dynamics. To the latter dynamics,
a power law for the relative order $m_{r}\sim -t^{-k}$ is found in the 
intermediate time-regime. The short-time scaling relations of small 
$m_{0}$ are generalized to an arbitrary $m_{0}$ and
all the time larger than $t_{mic}$. The characteristic functions
$\varphi(b,m_{0})$ for the scaling of $m_{0}$ 
and $\epsilon(b, {T'})$ for $T^{\prime}=T/T_{c}$ are obtained.
The crossover between scaling regimes is discussed in detail.

\end{abstract}

\newpage
\section*{1. Introduction}

In recent years, the universal scaling in non-equilibrium states have
attracted much attention. The phase-ordering process (POP) \cite{s3} 
and the short-time critical dyanmics (SCD) 
\cite{janss} are two fruitful examples.

In the critical dynamics, the 
short-time phenomena are those which happen at the
times just after a microscopic time-scale $t_{mic}$
needed for a system to forget its microscopic details, and much smaller than 
the macroscopic time scale $t_{mac}\sim \tau^{-\nu z}$. 
Since the pioneer work of H.K. Janssen et al \cite{s1}, universal 
short-time scaling has been found in a 
variety of different models (for a recent review, see \cite{s2}). 
It is believed that the singularity of the temporal 
correlation is essential to the short-time scaling. Therefore 
the scaling can emerge in the short-time regime of the evolution eventhough
all spatial correlations are still short-ranged. 

The dynamics to be considered here
has no conservation law, which is sometimes called model A \cite{s7}.
All critical exponents as well as the critical point can be fixed in 
the short-time regime \cite{batr92, li95, sch96}. This is important especially 
for Monte Carlo simulations because at 
the critical temperature a system needs infinite time 
to relax to equilibrium due to the critical slowing down. It
is good to avoid doing measurements in equilibrium states.
Therefore, it is interesting to investigate the connection between 
the short-time regime and the long-time regime and the crossover of
scaling patterns from one regime to another.

The power law increase of order in the SCD 
is valid only in the case of a small initial order-parameter $m_{0}$.
The general case of an arbitrary $m_{0}$ has been discussed
in \cite{s9}. 
The initial state with non-zero $m_{0}$ is off-critical. A characteristic
function must be introduced to describe the moving of $m_{0}$ in the
scale transformation. With
the characteristic function, the scaling law can
be extended to a large $m_{0}$ and all $t\gg t_{mic}$. 
As far as we know, only the numerical 
evidence has been found for this characteristic function. In this paper, the
soluble kinetic spherical model (KSM) will provide a concrete example.

Instead of a quench to the critical temperature $T_{c}$ in  the 
SCD, the system in the POP is quenched to a temperature 
lower than the critical one. 
For a system quenched from a symmetric
initial state, the final equilibrium is never achieved 
since the ergodicity is broken in the thermodynamics. Instead, the length-scale
of ordered regions grows with time as the different broken-symmetry
phases compete to select the equilibrium state.
This temporal singularity should also be the essence of the ordering
scaling. It has been found both experimentally 
and theoretically  \cite{bray99}
that the POP
at late times can be described by a single 
characteristic length-scale $L(t)\sim
t^{\rho}$, reflecting the self-similarity of domain patterns at different times. 

If the initial state breaks the symmetry, as the case of $m_{0}\neq 0$ to be 
discussed here, the equilibrium can be achieved. A natural question 
is how long the phase-ordering scaling will last. 
The non-zero $m_{0}$
generates a time-scale $t_{i}$ which characterizes the most interesting
short-time regime where we will see new scaling patterns.
Since $t_{i}\rightarrow \infty$ as $m_{0}\rightarrow 0$, the 
so-called short-time regime in fact can persist to a very long time.

The SCD and the POP are governed by two different fixed points, i.e., 
$T_{c}$ and $T=0$ respectively, therefore they have different scaling laws. 
It will
be interesting to see how the POP crosses over to the SCD as $T$ 
grows from $0$ to $T_{c}$. 
In the context of POP, besides $t_{i}$ there is another 
time-scale $t_{\mu}$ given by the non-zero temperature.
In this paper, we will introduce a characteristic
function to describe the temperature dependence of the process.
Since the critical exponents can not change smoothly, there must be a 
crossover region that separates the SCD scaling domain from the scaling 
domain of
POP. We will estimate the domain boundary through 
a self-consistency of the solution.

To obtain explicit analytical results, 
we confine ourself to the soluble KSM
with or without long-ranged interactions. 
The KSM is a generalization to the 
static spherical model \cite{s4}, to which a model A dynamics is given
by the Langevin equation. 
The static spherical model has been proven to be equivalent
to the large-$n$ limit $n$-vector model (LLNM) in static properties \cite{s5}. 
When the dynamics is included, they should be still equivalent in the
thermodynamic limit since the fluctuation of $<\phi^{2}>$ in the LLNM
can be neglected. The LLNM being a soluble kinetic model with a 
dimensionality higher than one has been extensively studied by many authors.
The SCD of the LLNM with the short-ranged interaction
and small $m_{0}$ is well-known \cite{s1, s15, s15b}. 
To the symmetric POP, i.e., $m_{0}=0$, many results have
been obtained for the model with either the short-ranged or the long-ranged
interaction \cite{s17, s21, s20, s19}. 
We will use the language of KSM to present our new results originated from
the non-symmetry initial state, i.e., 
an arbitrary $m_{0}$. The KSM has the domain-wall picture which is
helpful for understanding the ordering process and the crossover from the
POP to the SCD. 

In this paper, we will concentrate on the consequences of non-symmetric
initial states. The time-dependent
order parameter $m(t)$ 
as well as the response propagator and correlation function are
calculated. 
The dynamic behavior of the symmetric KSM with short-ranged 
interaction in the long-time regime 
was already studied in  \cite{s8}, where no trace of the initial condition
is left. 
In the case of non-zero $m_{0}$,
the order-parameter has
non-equilibrium values in the short-time regime after the system 
is quenched from a very high temperature.
It will show rich scaling patterns governed by the double 
time-scales $t_{i}$ and $t_{\mu}$.

It has been known that the critical exponents can be modified by
the interaction range. How the characteristic critical exponent $\theta$
or ${\theta'}$ depends on the interaction range has not been explicitly
shown. But from the general consideration of that the longer interaction range
and higher dimension are both in favor of the spacial correlation, the long
interaction range will decrease the exponent ${\theta'}$. We will show that
is just the case in the KSM.
Particularly we show  that 
the interaction-range parameter $\sigma$
and the dimensionality $d$ enter the time-dependent order-parameter 
$m(t)$ always in the combination $d/\sigma$.

The remainder of this paper is organized as follows.
In section 2 the kinetic spherical model 
is introduced. Laplace transformation to $m(t)$
 is established in section 3. Solutions and
scaling behaviors
at $T=T_{c}$ and $T<T_{c}$ are obtained in section 4
and section 5, respectively. In section 6, a concrete example of 
$d/ \sigma=3/2$ is given. Our conclusions and discussions will be given 
in the last section.

\section*{2. The Model}

\par
   The Hamiltonian of the spherical model is
   \begin{equation}
   \label{ee1}
   H={\frac{\alpha}{2}}\sum\limits_{i} S ^{2}_{i}-
   {\frac{\beta}{2}}\sum\limits_{ij}J_{ij}S_{i}S_{j}
   \end{equation}
with the constraint
   \begin{equation}
   \label{ee2}                                                      
   {\sum\limits_{i} S ^{2}_{i}}=N
   \end{equation}
where $i, j$ are labels of lattice sites, $N$ is the total number of spins;
$\beta={\frac {1}{k_{B} T}}$. 
In the dynamic process, $\alpha$ is a
time-dependent Lagrange multiplier corresponding to the
constraint.
Joyce \cite{s6} first studied the static spherical model
with long-ranged ferromagnetic interactions. In a $d$-dimensional lattice,
$$J_{ij}={J_{0}r^{-(d+s)}_{ij}/ \sum\limits_{j}r^{-(d+s)}_{ij}}$$
with $0<s<2$ for long-ranged interactions, while $s > 2$
for short-ranged interactions. Where $r_{ij}$ is the distance between
the sites $i$ and $j$.

\par
The Langevin equation for this constrained spin system is
   \begin{equation}
   \label{ee3}
   {\frac{\partial S_{i}}{\partial t}} 
   =-\lambda\alpha S_{i}+\lambda\beta\sum\limits_{j}J_{ij}S_{j}+
   \eta_{i} 
   \end{equation} 
where $\lambda$ is the kinetic coefficient and $\eta_{i}$
being a Gaussian white noise characterized by
   \begin{equation}
   \label{ee4}
   < \eta_{i} (t) >=0,\quad <\eta_{i} (t)\eta_{j} (t^{\prime}) >=2\lambda
   \delta_{ij}\delta (t-t^{\prime} )
   \end{equation}
where angle brackets mean average over the noise.
\par
Since the constraint is applied to the whole
dynamic process, at any time
$t$, there is a consistency condition
    \begin{equation}
    \label{ee5}
    <\sum\limits_{i} S_{i} {\frac{\partial S_{i}}{\partial t}}>=0
    \end{equation}

\par
The time-dependent order-parameter
$m(t)={\frac{1}{N}} <\sum\limits_{i} S_{i}>$
is in the translationally invariant case  equal to
$<S_{i}>$. Taking the average over the noise in  (\ref{ee3}), one has
   \begin{equation}
   \label{ee6}
   {\frac{\partial m(t)}{\partial t}}=-\lambda\tau (t)m(t)
   \end{equation}
where $\tau (t)=\alpha (t) -\beta J_{0}$.
\par
The fluctuations of spins are defined to be
  $ \widetilde S _{i}=S_{i}-m(t)$.
In the momentum space, one has
$$\widetilde S (p,t)={\frac{1}{\sqrt{N}}}\sum\limits_{i}
\widetilde S_{i} e^{i {\bf p}\cdot {\bf r_{i}}}, \quad
\eta (p,t)={\frac{1}{\sqrt{N}}}\sum\limits_{i}
\eta_{i} e^{i {\bf p}\cdot {\bf r_{i}}}.$$
with $r_{i}$ the position vector of site $i$, and 
$$J(p)=\sum\limits_{j} J_{ij} e^{i{\bf p}\cdot 
{\bf r_{ij} }}$$
with $J(0)=J_{0}$.
Corresponding to (\ref{ee3}), 
the dynamic equation for $\widetilde S (p,t)$ is
   \begin{equation}
   \label{ee7}
   {\frac{\partial \widetilde S (p,t)}{\partial t}}=-\lambda
   (\tau (t)+\Delta (p))\widetilde S (p,t)+\eta (p,t) 
   \end{equation}
where $\Delta (p)=\beta (J_{0}-J(p))$. The consistency condition
gives
   \begin{eqnarray}
   \label{ee8}
   \tau (t) & = & \tau_{sub}+\beta J_{0}\left [ m^{2}
   (t)-1\right ] +
   {\frac{\beta }{N}}
   \sum\limits_{p} J(p) < \widetilde S (-p,t)
   \widetilde S(p,t)> \nonumber \\
   & + &
   {\frac{1}{\lambda N}}
   \sum\limits_{p}< \widetilde S (-p,t)\eta (p,t)> 
   \end{eqnarray} 
where the first term comes from the mass subtraction which 
guarantees $\tau_{c}(\infty)=0$ at the
critical point.
It can be shown that $\tau_{sub}=1$.
Due to causality, the last term of (8) is zero.
\par
By solving (7), it is not difficult
to obtain the response propagator
  \begin{equation}
  \label{ee9}
  G_{p}(t,t^{\prime})={\frac {1}{2\lambda}}
  < \widetilde S (-p,t)\eta (p,t^{\prime})>
  =\Theta (t-t^{\prime})e^{-\lambda\Delta (p)
  (t-t^{\prime})-\lambda\int\limits_{t^{\prime}
  }^{t}dt^{\prime\prime}
  \tau (t^{\prime\prime})}  
  \end{equation}
with the Heaviside step function 
$ \Theta (t-t^{\prime})= 1 $ for $t>t^{\prime}$, otherwise
$\Theta (t-t^{\prime})=0$;
and the full correlation function (correlation function including
the initial correlation)
  \begin{eqnarray}
  \label{ee10}
  \widetilde C_{p}(t,t') & = &
  < \widetilde S (p,t)\widetilde S (-p,t^{\prime})> \nonumber \\
  & = &
  < \widetilde S (p,0)\widetilde S (-p,0)>
  G_{p}(t,0)G_{-p}(t^{\prime},0)+C_{p}(t,t^{\prime})
  \end{eqnarray}
with the correlation function 
  \begin{equation}
  \label{ee11}
  C_{p}(t,t^{\prime})=2\lambda\int\limits_{0}^{\infty}
  dt^{\prime\prime}G_{p}(t,t^{\prime\prime})
  G_{p}(t^{\prime},t^{\prime\prime})
  \end{equation}
where $<\widetilde S (p,0)\widetilde S (-p,0)>$ is given by
the initial state. Consider a system initially at a
temperature $T \gg T_{c}$ with a given initial order-parameter
$m_{0}$. Since initial
correlations are short-ranged, one has
$<\widetilde S (p,0)\widetilde S (-p,0)>=
(1 - m^{2}_{0})$.

\section*{3. Laplace transformation}
\par
Introducing $ f(t)=m^{-2} (t)$, from (6) and (8) one obtains a linear
integrodifferential equation for $f(t)$
    \begin{equation}
    \label{ee12}
    {\frac {\partial f(t)} {\partial t}}=2\lambda\beta J_{0}
    -2\lambda( \beta J_{0} -1)f(t)+{\frac{2\lambda\beta }{N}}
    \sum\limits_{p} J(p) \widetilde C_{p}(t,t)f(t)
    \end{equation}

For a $d$-dimensional cubic lattice,  
$$J(p)=J_{0}C_{d,s}(p)/C_{d,s}(0)$$
with $C_{d,s}(p)=\sum\limits_{l}
|{\bf l} |^{-(d+s)} \cos( {\bf l}\cdot {\bf p}) $
and ${\bf l}\cdot {\bf p}=l_{1}p_{1}+...+l_{d}p_{d}$.
The critical temperature is given by \cite{s6}
       \begin{equation}
       \label{ee13}
 \beta_{c}J_{0}={\frac {1}{N}}\sum\limits_{p}
[1-C_{d,s}(p)/C_{d,s}(0)]^{-1}
        \end{equation}
Define
    \begin{equation}
    \label{ee14}
    w(x)={\frac{1}{N}}\sum\limits_{p} {\frac{1}{x+
    \Delta (p)/(\beta J_{0})}}
    \end{equation}
then $\beta_{c} J_{0} =w(0)$. By Laplace transformation 

    $$F(q)=\int\limits_{0}^{\infty}  dt f(t) e^{-qt}, $$
Equation (12) is transformed to
        \begin{equation}
        \label{ee15}
   F(q)={\frac{
    {\frac{\beta J_{0}}{q}} +{\frac{1}{2\lambda}}
    (m_{0}^{-2}-1)w({\frac{q}{2\lambda \beta J_{0}}})}
    {\beta J_{0} -w({\frac{q}{2\lambda \beta J_{0}}})}}
     \end{equation}
Using properties of Laplace-transformation, it
is easy to obtain
    \begin{equation}
    \label{ee16}
    \lim_{q \rightarrow \infty} q F(q)=m_{0}^{-2}
    \end{equation}
    \begin{equation}
    \label{ee17}
    \lim_{q \rightarrow 0} q F(q)=m^{-2}(\infty)=(1-
    T/T_{c})^{-1}
    \end{equation}
Hence the initial condition is satisfied, and the infinite
time limit correctly recovers the equilibrium result.
From (17), one can also extract the critical exponent $\beta
=1/2$.

For small ${p}$, one has
    \begin{equation}
    \label{ee18}
    1-C_{d,s}(p)/C_{d,s}(0)=Cp^{\sigma}
    \end{equation}
where the constant $C$ depends only on $d$ and $s$. The parameter $\sigma$
denotes an effective potential-range defined by \cite{s10}, with
$\sigma=s$  for $0<s<2$  and  $\sigma=2$ for $s>2$.

From here after, we suppose $d/2<\sigma<d$. For the infinite 
system, the sum in (14) is replaced by an integral.
Combining (13) with (18), in the continuum-limit one has  
   \begin{equation}
   \label{ee19}
   w(x)=w(0)-Dx^{d/\sigma-1}
   \end{equation}
where the constant $D$ again depends only on $\sigma$ and $d$ and is positive. Then (15) becomes
    \begin{equation}
    \label{ee20}
    F(q)={\frac{
    {\frac{\beta J_{0}}{q}} +{\frac{\beta J_{0}}{2\lambda}}
    (m_{0}^{-2}-1)}
    {(\beta -\beta_{c})) J_{0}+D({\frac{q}{2\lambda \beta J_{0}}}
    )^{d/\sigma-1}}}-{\frac{1}{2\lambda}}(m_{0}^{-2}-1)
    \end{equation}

Since the microscopic details have been skipped in the continuum
limit,  (20) does not give the correct initial value.
Instead, it gives $f(0)=\infty$ corresponding to $m(0)=0$.
Therefore  (20) is only valid for $t$ larger than a 
microscopic time-scale $t_{mic}$. The last term of  (20) only has 
contribution at $t=0$, therefore it can be dropped for the regime 
$t>t_{mic}$.

Notably, if one rescales the kinetic coefficient $\lambda$
and the initial order $m_{0}$ as following
    \begin{equation}
    \label{ee21}
    \lambda^{\prime}=\lambda D_{c}^{{\frac{1}{1-d/\sigma}}}
    \end{equation}
    \begin{equation}
    \label{ee22}
    m_{0}^{\prime}=\left [ D_{c}^{{\frac{1}{1-d/\sigma}}} 
    (m_{0}^{-2}-1)+1\right ]^{-1/2}
    \end{equation}
with $D_{c}=D/(\beta_{c}J_{0})^{d/\sigma}$, 
then $F(q)$ depends on the dimension $d$ and the interaction-range
parameter $\sigma$ through their ratio $k=d/\sigma$, i.e., 
for $t>t_{mic}$,
    \begin{equation}
    \label{ee23}
    F(q)={\frac{
    {1 \over q} +{1 \over 2\lambda^{\prime}}
    ({m'}_{0}^{-2}-1)}
    {\mu+(1-\mu)^{k}({q \over 2\lambda^{\prime}})^{k-1}}}
    \end{equation}
where $\mu=(T_{c}-T)/T_{c}$ is the reduced temperature. $D_{c}$ is a
non-universal constant depending on $d$, $\sigma$ and the regularization
scheme. As one will see in the next section, (22) follows from a rescaling of 
time by a factor $D_{c}^{{1\over 1-k}}$. 

 Equation (23) is the starting point of the following discussions.

\section*{4. $T=T_{c}$}
\par
At the critical point, $\mu=0$, the Laplace back-transformation
of  (\ref{ee23}) 
is easily carried out. The solution for the order parameter at
the critical point is 
    \begin{equation}
    \label{ee24}
    m(t)= \pm \left [{\frac {\Gamma (k)}
    {(2\lambda^{\prime})^{k-1}t_{i}}}\right ]^{1/2}
    (1+t/t_{i})^{-1/2}t^{1-{k\over 2}}
    \end{equation}
where $t_{i}$ is a characteristic time-scale of the short-time regime, 
which depends on the initial condition,
   \begin{equation}
   \label{ee25}
   t_{i}={\frac {(k-1)(m^{-2}_{0}-1)}
   {2\lambda }}
   ={\frac {(k-1)({m'}^{-2}_{0}-1)}
   {2\lambda^{\prime} }}
   \end{equation}
It is invariant in the combined transformation of (\ref{ee21}) and (\ref{ee22}).  
The sign of $m(t)$ can be found from the symmetry of the
initial state. The $|m(t)|$ firstly increases then it relaxes to
zero. The maximum is at $t={2-k \over k-1}t_{i}$, it is propotional to $t_{i}$.

For small enough $m_{0}$, the $t_{i}$ could be much larger than $t_{mic}$.
Then in the regime $t_{mic}<t\ll t_{i}$ one can observe 
the power-law order increasing 
suggested by Janssen et al \cite{s1},
    \begin{equation}
    \label{ee27}
    m(t)\sim t^{\theta^{\prime}}
    \end{equation}
with the exponent  $  \theta^{\prime}=1-{k \over 2}$. This regime is called 
the critical initial slip. A standard explanation to the increase
of the order
is that the order-parameter follows a mean-field ordering process because
$T_{c}<T_{c}^{m.f.}$ as long as correlations are short-ranged in the 
critical initial slip. However the exponent $\theta^{\prime}$ is non-trivial.
As we will see in the next section, 
it is different from the corresponding exponent for 
the POP. 
    
In the long-time regime, $t \gg t_{i}$,
    \begin{equation}
    \label{ee28}
    m(t)=\left [{\Gamma (k)\over
     (2\lambda^{\prime})^{k-1}}\right ]^{1/2}
    t^{-{k-1 \over 2}}
    \end{equation}
The dependence on the initial condition disappeares as expected.
Comparing with the well-known nonlinear relaxation scaling law, one
obtains ${\beta \over \nu z}={k-1 \over 2}$.

When $m_{0}\neq 0$, the initial state is not at the fixed point. 
However,
as proposed by Zheng \cite{s9}, the short-time critical scaling
hypothesis could be generalized to an arbitrary $m_{0}$ by 
introducing a characteristic function of $m_{0}$ and the scaling 
factor $b$. 

As the length is rescaled by a factor $b$, supposing $t_{i}$ is rescaled
by $b^{-z}$ just like a time, then $m(t)$ given by (\ref{ee24}) satisfies  
the following scaling relation
\begin{equation}
\label{ee29}
m(t,m_{0})=b^{-{\beta \over \nu}} m(b^{-z}t, \varphi(b, m_{0}))
\end{equation}
where $\varphi(b, m_{0})$ is the characteristic function, which describes 
how the initial order changes in the scale transformation.
Its explicit
form can be deduced from the scaling of $t_{i}$
\begin{equation}
\label{ee30}
\varphi(b,m_{0})=[b^{-z}(m^{-2}_{0}-1)+1]^{-{1\over 2}}
\end{equation}
The effective dimension of the initial order $m_{0}$ is defined by 
\begin{equation}
\label{ee31}
x_{0}(b,m_{0})={\ln[\varphi(b,m_{0})/m_{0}] \over \ln b}
\end{equation}
As expected, as $m_{0}\rightarrow 0$, $\varphi(b,m_{0})=b^{z/2}m_{0}$,
so that the initial order has an anomalous dimension
$x_{0}={z \over 2}$ and the power law (\ref{ee27}) is
recovered. When $m_{0}$ takes its maximum $m_{0}=1$, $\varphi=1$ and
$x_{0}=0$. In this case, $t_{i}=0$, hence only the nonlinear relaxation
scaling law, $m(t)\sim t^{-{\beta \over \nu z}}$, is observed. Figure 1
plots $x_{0}$ versus $m_{0}$ for $b=2$ and $b=4$. It is qualitatively
consistent with the numerical results of Ising model \cite{s2}
, for which $x_{0}(b, m_{0})$ is a monotonously decreasing
function of $m_{0}$. This appeares to be reasonable
 by recalling that the spherical model is just a very rough
approximation of the Ising model.

Substituting (24) into (6), one gets
$\tau (t)={\frac {k-2}{2\lambda t}}+{\frac {1}{2\lambda
(t+t_{i})}}$. This result together with (9) gives the response
propagator
    \begin{equation}
    \label{ee32}
    G_{p}(t,t^{\prime})=\Theta (t-t^{\prime})\; ({\frac {t}
    {t^{\prime}}})^{ (2-k)/2
    } \; ({\frac {t^{\prime}+t_{i}}{t+t_{i}}})^{
    1/2} \; e^{{-\widetilde\lambda p^{\sigma}(t-t^{\prime})}}
    \end{equation}
where $\widetilde \lambda=\lambda C \beta J_{0}$ with $\beta=\beta_{c}$
in this case. The correlation function  (11)
is given by
    \begin{eqnarray}
    \label{ee33}
    C_{p}(t,t^{\prime}) & = & 2\lambda t^{\prime}
    ({\frac {t}{t^{\prime}}})^{
    {\frac {2-k}{2}}} ({\frac {t^{\prime}+t_{i}}
    {t+t_{i}}})^{{\frac {1}{2}}}
    e^{{-\widetilde\lambda 
    p^{\sigma}(t-t^{\prime})}} \nonumber \\
    &  & \cdot \int\limits_{0}^{1}dy(1-y)^{k-2}{\frac{t^{\prime}-
    t^{\prime}y+t_{i}}{t^{\prime}+t_{i}}
    } e^{-2\widetilde\lambda p^{\sigma}t^{\prime}y}
    \end{eqnarray}

When $t\ll t_{i}$ and $t^{\prime}\ll t_{i}$,  (\ref{ee32}) has a scaling form,
$$G_{p}(t,t^{\prime})=p^{-2+\eta +z}
({\frac {t}{t^{\prime}}})^{\theta} h(p^{z} (t-t^{\prime}))$$
with $h(x)$ being a scaling function, 
as suggested in \cite{janss}. The exponents $\theta=1-{k \over 2}$, $z=\sigma$
and  $\eta=2-\sigma$ can be read out. The values of $z$ and $\eta$ agree
with the corresponding results in \cite{s101,s13, s13b} and \cite{s6,s11,s12}.
Two exponents $\theta$ and $\theta^{\prime}$ are equal to each other.
For short-ranged interactions $\sigma=2$, they coincide with the 
results of LLNM \cite{s1,s14,s15, s15b}.  

When $t=t^{\prime}\ll t_{i}$,
we get the equal-time correlation function
    \begin{equation}
    \label{ee34}
    C_{p}(t,t)={2\lambda t e^{-2\widetilde\lambda p^{\sigma}t} \over
    k-1} (1+O(2\widetilde\lambda 
    p^{\sigma} t))
    \end{equation}
It has the scaling form \cite{s1,s15, s15b,s17,s16}
$C_{p}(t,t)=p^{-2+\eta}g(p\xi_{c}(t))$
with $g(x)$ being another scaling function. 
Where $\xi_{c}(t)\sim t^{1/\sigma}$ is a characteristic length-scale. 
Physically, it should be related to the correlation length.

When $t,t^{\prime}\gg t_{i}$ and 
$2\widetilde\lambda p^{\sigma}t^{\prime}$ large,
the propagator and equal-time
correlation function reads as
    \begin{equation}
    \label{ee35}
    G_{p}(t,t^{\prime})=\Theta (t-t^{\prime})
    ({\frac {t}{t^{\prime}}}
    )^{-{\frac {k-1}{2}}}
    e^{-\widetilde\lambda p^{\sigma}(t-t^{\prime})}
    \end{equation}
    \begin{equation}
      \label{ee36}
    C_{p}(t,t)={p^{-\sigma} \over C\beta_{c} J_{0}}
    \left [1-{\frac {k-1}
    {2\widetilde\lambda p^{\sigma}t}
    }+O({\frac {1}{(\widetilde\lambda p^{\sigma}t)^{2}}})\right ]
    \end{equation}

By use of the characteristic function  (\ref{ee30}), the response propagator and 
correlation function have general scaling forms which are valid for the
times up to $t_{mic}$ and for an arbitrary $m_{0}$,
\begin{equation}
  \label{ee37}
G_{p}(t,t',m_{0})=p^{-2+\eta+z}\widetilde h(p\xi_{c}(t),
p\xi_{c}(t'),\varphi (p^{-1},m_{0}))
\end{equation}
\begin{equation}
  \label{ee38}
C_{p}(t,t',m_{0})=p^{-2+\eta}\widetilde g(p\xi_{c}(t),p\xi_{c}(t'),
\varphi (p^{-1},m_{0}))
\end{equation}
where $\widetilde h$ and $\widetilde g$ are scaling functions. 
The significance of the characteristic function is to connect the 
short-time regime and the long-time regime so that 
for each thermodynamic
quantity only one scaling function is needed for the whole physical interesting
time regime.

\section*{ 5. $T<T_{c}$}

In the case of $T<T_{c}$, the system undergoes an ordering process
which is governed by the fixed point $T=0$. For large $\beta$, 
i.e., $\mu\rightarrow 1$, one can expand the denominator of  (23) to get 
    \begin{equation}
      \label{ee39}
    F(q)=\mu^{-1}\left [{1 \over q}-
    {2\lambda^{\prime}\over \mu}({1-\mu \over 2\lambda^{\prime} 
    })^{k}q^{k-2}-{{m'}_{0}^{-2}-1\over \mu}({1-\mu \over
    2\lambda^{\prime}})^{k} q^{k-1}\right ]
    \end{equation}
The asymptotic solution for the bulk order
is obtained by the Laplace back-transformation
    \begin{equation}
      \label{ee40}
    m(t)=m(\infty)[1-
    (1-{\frac {t_{i}}{t}})
    ({\frac {t}{t_{\mu}}})^{1-k}]^
    {-1/2}
    \end{equation}
where  $t_{\mu}=[
{2\lambda^{\prime} \over |\mu| \Gamma(2-k)}({1-\mu \over 2
\lambda^{\prime}})^{k}]^{{1 \over k-1}}$. Again for a fixed $k$, the dependence on $d $ and $\sigma$ 
can be absorbed by a proper rescaling of the time.

The $t_{\mu}$ is a temperature-dependent
time-scale. The corresponding length-scale should be the domain-width
$\xi_{dw}\sim t_{\mu}^{1/z}$. The POP scaling only could survive
when $\xi_{dw}$ is smaller than the domain-size which grows with time as
$\xi \sim t^{1/z}$. That implies $t_{\mu}$ should be smaller than a 
characteristic time of the phase-ordering regime. From the  
consistency of  (\ref{ee40}) we do find a constraint for $t_{\mu}$
$${t_{\mu} \over t_{i}} <{k^{{k \over k-1}} \over  k-1}.$$
Replacing $t_{\mu}$ by its explicit expression, one has
   \begin{equation}
     \label{ee41}
   (1-{T \over T_{c}})({T_{c}\over T})^{k}>
   {1\over k\Gamma(2-k)}(
   {k-1 \over 2\lambda^{\prime} k  t_{i}})^{k-1}
   \end{equation}
The r.h.s is a positive constant. The l.h.s is a 
monotonously decreasing function with respect to $T$, which has 
zero value at $T=T_{c}$. Hence 
there is a boundary temperature $T_{b}<T_{c}$. The ordering scaling appears 
in the regime $0<T<T_{b}$. The smaller $m_{0}$, the larger $T_{b}$. It tends to
$T_{c}$ as $m_{0}\rightarrow 0$.

As $t\ll t_{i}$, one has
\begin{equation}
  \label{ee42}
m(t)=m(\infty )\left [1+({\frac {t^{\prime}_{i}}{t}})^{k}
\right ]^{-{\frac {1}{2}}}
\end{equation}
where $t_{i}^{\prime}=t_{\mu}(t_{i}/t_{\mu})^{1/k}$ is a
time-scale that characterizes the
short-time ordering behavior. 
In the symmetry case, $t_{i}^{\prime}=\infty$ so $m(t)=0$. 
No information about the POP can be
extracted from the bulk order-parameter. 
With a 
non-symmetric initial state however, 
the  process has various scaling behaviors 
manifested by $m(t)$ due to the competition of two time-scales.
From the self-consistency condition mentioned above, one can show that
in the  scaling regime, $t_{i}$ is at least larger than
$t_{i}^{\prime}/2$. There will be three different scaling regimes.

(a) $t_{mic}<t\ll {t'}_{i}$ and $t_{i}$

This is the
extensively studied regime in the symmetric POP.
In literatures, sometimes it is called the late time stage. 
There is no contradiction because in the symmetry case $t_{i}$ and $t_{i}^
{\prime}$ are infinite.  
The late-time simply means $t>t_{mic}$. 

For small $m_{0}$,
${t'}_{i}$ and $t_{i}$ can be much larger than $t_{mic}$. 
Then in this regime, 
one can observe a power law increase of the order-parameter
    \begin{equation}
      \label{ee43}
    m(t)=m(\infty)({\frac {t}{t^{\prime}_{i}}})^{\theta_{T}^{\prime}}
    \end{equation}
with $\theta_{T}^{\prime}=k/2$. This is the initial increase in the
POP. 

The response propagator
when $t\ll {t'}_{i}$ and $t_{i}$ is
    \begin{equation}
      \label{ee44}
    G_{p}(t,t^{\prime})=\Theta (t-t^{\prime}) ({\frac {t}
    {t^{\prime}}})^{
    \theta_{T}} e^{{-\widetilde\lambda
    p^{\sigma}(t-t^{\prime})}}
    \end{equation}
with $ \theta_{T}=k/2$. 
We note that \cite{s16} got the exponent $\theta_{T}$ at $\sigma=2$
for the response propagator in the study of the domain growth.

(b)  $t^{\prime}_{i}\ll t\ll t_{i}$

The existence of this regime depends on both the temperature $T$ and the
initial order $m_{0}$. The self-consistency of  (\ref{ee40}) implies 
${t^{\prime}_{i} \over t_{i}} < k (k-1)^{(1-k)/k}$. As $T\rightarrow 0$,
${t'}_{i}$ tends to zero while $t_{i}$ does not change. On  the other hand, 
as far as $k>1$, $t_{i}$ tends to infinite faster than ${t'}_{i}$ as $m_{0}
\rightarrow 0$. Therefore, this regime exists if $m_{0}$ is small but not
exactly zero and the temperature is low.

Define a relative order with respect to the equilibrium order,

$$ m_{r}(t, {T'}, m_{0})={m(t)-m(\infty)\over m(\infty)} $$
where ${T'}=T/T_{c}$. One can easily obtain from (\ref{ee43})
    \begin{equation}
      \label{ee45}
   m_{r}(t)= -{\frac {1}{2}}
    ({\frac {t^{\prime}_{i}}{t}})^{k}
    \end{equation}
The $m_{r}$ asymptotically follows
a power law. 

When ${t'}_{i}\ll t \ll t_{i}$ while ${t'}\ll {t'}_{i}$, the response 
propagator is
    \begin{equation}
      \label{ee46}
    G_{p}(t,t^{\prime})=\Theta (t-t^{\prime}) ({\frac {{t'}_{i}}
    {t^{\prime}}})^{
    \theta_{T}} e^{{-\widetilde\lambda
    p^{\sigma}(t-t^{\prime})}}
    \end{equation}
The time-scale $t^{\prime}_{i}$  enters the propagator.

Another case is ${t'}_{i}\ll {t'}, t \ll t_{i}$. The response
propagator has an asymptotic form
    \begin{equation}
      \label{ee47}
    G_{p}(t,t^{\prime})=\Theta (t-t^{\prime})e^{{-\widetilde\lambda
    p^{\sigma}(t-t^{\prime})}}
    \end{equation}
The corresponding equal-time correlation function is
    \begin{equation}
      \label{ee48}
    C_{p}(t,t)={\lambda \over \widetilde\lambda p^{\sigma}}
    (1-e^{{-2\widetilde\lambda
    p^{\sigma}t}})
    \end{equation}
It exponentially tends to the structure function of the equilibrium state.

(c) $t\gg t_{i}$

Notably, in the long-time regime (including
all time if $m_{0}=1$)
, $|m(t)|$ approaches $|m(\infty)|$ from above if $m_{0}\neq 0$, and
   \begin{equation}
     \label{ee49}
   m_{r}(t) 
   = {\frac {1}{2}}({\frac{t_{\mu}}{t}})^{k-1}
   \end{equation}
The asymptotic power law has an
exponent $\theta^{\prime}_{1}=1-k$ , which is different from
$-\beta/\nu z={\frac {1-k}{2}}$ in the critical
nonlinear relaxation. This is in agreement with the results of 
\cite{s8,s18} in the special case of $\sigma=2$. 
It can be understood since two processes are governed by 
different fixed points. 

In the general case, $0<T<T_{b}$ and $m_{0}\neq 0$, 
the dimension of the
order-parameter is scale-dependent.
As one has seen in the cases (b) and (c),
a more suitable scaling operator is the relative order $m_{r}$.
There exist two
off-critical parameters, ${T'}$ and $m_{0}$.
They will be running versus the scale in a scale transformation. 
Therefore one needs two characteristic
functions. The characteristic function for $m_{0}$ has been 
given in (\ref{ee30}). The charateristic function
for ${T'}$ can be obtained by requiring $t_{\mu}$
scaling as $t$, then one can have a general scaling relation,
\begin{equation}
  \label{ee50}
m_{r}(t, {T'}, m_{0})=m_{r}(b^{-z}t, \epsilon(b,{T'}),\varphi(b,m_{0}))
\end{equation}
The characteristic function $\epsilon(b,{T'})$ satisfies the following equation
\begin{equation}
  \label{ee51}
{\epsilon^{k}(b, {T'})
\over 1-\epsilon(b, {T'})}=b^{-z(k-1)}{{T'}^{k} \over 1-{T'}}
\end{equation}
It can be easily checked that $\epsilon(b, 0)=0$, $\epsilon(b, 1)=1$,
$\epsilon(1, {T'})=1$, and $\epsilon(\infty, {T'})=0$. In fact (\ref{ee50}) 
is valid for all
temperature (if $m(\infty)$ is replaced by $\sqrt{|\mu|}$ when $T>T_{c}$) 
as one will see from the example of the next section. 
One can again define the effective dimension of ${T'}$ as
$$y(b, {T'})={\ln(\epsilon(b, {T'})/{T'}) \over \ln({T'})}$$
When $T\rightarrow 0$,
one has a constant dimension for ${T'}$, i.e., $y(b,0)=-z{k-1\over k}$. 
It is negative since $T=0$ is an attractive fixed point. For $k=3/2$ and
 $b=2, 4$, $y(b,
{T'})$ is plotted versus ${T'}$ in Figure 2.
In the vicinity
of the critical temperature, the scaling of the reduced temperature can
be recovered from (\ref{ee51}), i.e., 
$\mu(b)=1-\epsilon(b,T_{c})=b^{z(k-1)}\mu$. The critical exponent
$\nu=z(k-1)$ is obtained. 

As in the critical dynamics, with the help of the characteristic functions, the
response propagator and correlation function can be casted into 
scaling functions
\begin{equation}
  \label{ee52}
G_{p}(t,t',{T'},m_{0})=\widetilde h^{\prime}\left (p\xi (t),
p\xi (t'),\epsilon(p^{-1},{T'}),\varphi (p^{-1},m_{0})\right )
\end{equation}
\begin{equation}
  \label{ee53}
C_{p}(t,t',{T'},m_{0})=p^{-\sigma}\widetilde g^{\prime}\left (p\xi (t),
p\xi (t'),\epsilon(p^{-1},{T'}),\varphi (p^{-1},m_{0})\right )
\end{equation}
Where the characteristic length $\xi(t) \sim t^{\rho}$ with $\rho={1\over
\sigma}$
is the
domain size \cite{s3}.
Apart from some arbitrary coefficients, $\widetilde h^{\prime}$ and
$\widetilde g^{\prime}$ are universal.

\section*{ 6. Example of $k=3/2$}

In order to see the crossover more clearly, let us 
consider the concrete example of $k=3/2$.  The 3-dimensional model with
the short-ranged interaction belongs to this case.
A compact solution for m(t) is available. 

(a) $T\le T_{c}$

A direct Laplace back-transformation to  (23) gives
    \begin{equation}
      \label{ee54}
    m(t)=m(\infty)\left [1+{2t_{i} \over \pi t_{\mu}}\sqrt{{t_{\mu} 
    \over t}}
    -(1+{2t_{i} \over \pi t_{\mu}})
    e^{{t \over \pi t_{\mu}}} 
    erfc(\sqrt{{t \over \pi t_{\mu}}})\right ]^{-1/2}
    \end{equation}
where
$erfc({x})={\frac {2}{\sqrt{\pi}}}\int\limits_{x}^{
\infty} e^{-\tau^{2}} d\tau$ is the complementary error
function.

Taking the limit $\mu=0$ ($T=T_{c}$), 
the critical scaling of  (24) for 
$k=3/2$ is recovered. In the other hand, for small $T$, $\mu \sim 1$, 
one has the  ordering scaling for $k=3/2$ 
given in section 5.

Since no expansions around $T=0$ have been made in obtaining
$m(t)$,  (\ref{ee54}) is valid at temperatures equal and
lower than $T_{c}$.

In Figure 3, we compare the results of (\ref{ee40}) and (\ref{ee54}). As the
 self-consistency
condition (\ref{ee41}) implies, $m(t)$ of (\ref{ee40}) graduately deviates from the 
correct results of (\ref{ee54}) as the temperature increases, and becomes singular 
when it is bigger than certain temperature.

(b) $T>T_{c}$

For completeness, let us also write down the exponential relaxation
in the high temperature phase. Since $\beta_{c}>\beta$,   (23)
has a positive singular point $(\pi t_{\mu})^{-1}$,
\begin{equation}
  \label{ee55}
F(q)={1\over |\mu|(\pi t_{\mu})^{1/2}}{ 1+2 t_{i} q
\over q(q^{1/2}-(\pi t_{\mu})^{-1/2})}
\end{equation}
The Laplace back-transformation gives
\begin{equation}
  \label{ee56}
m(t)=\sqrt{|\mu|}\left [ 
{2t_{i} \over \pi t_{\mu}}\sqrt{{t_{\mu} \over t}}-1
    +(1+{2t_{i} \over \pi t_{\mu}})
    e^{{t \over \pi t_{\mu}}}\left (2- 
    erfc(\sqrt{{t \over \pi t_{\mu}}})\right )\right ]^{-{1\over 2}}
\end{equation}

In the long-time regime, one has the familiar exponential decay
$$m(t) \sim e^{-{t\over 2 \pi t_{\mu}}}$$

\section*{ 7. Conclusion and discussions}

In summary, we solve the KSM with long-ranged interactions. 
The system is quenched from a very high temperature into either  
$T=T_{c}$ or $T<T_{c}$, with an arbitrary $m_{0}$.
The bulk order as well as the response 
propagator and correlation function are calculated.

For small $m_{0}$, we recover the 
power law of the initial increase in the SCD. 
 We obtain
$\theta^{\prime}=\theta=1-{k\over 2}$. For the short-ranged interaction,
they agree with the existing results. 

The initial order increasing power law is also found in the 
POP, but with the exponent $\theta^{\prime}_{T}=k/2$ different
from that in the critical dynamics. As the symmetry is
broken, a
 new scaling regime emerges. In the time range
 ${t'}_{i}<t<t_{i}$  the relative
order has a power law $m_{r}\sim -t^{-k}$, while in the long-time regime
$m_{r}\sim t^{-(k-1)}$.

We obtain the characteristic function of $m_{0}$  
which enables us to generalize the critical short-time scaling relation
to the whole time range and to an arbitrary $m_{0}$. 

As in the SCD, the whole ordering process with broken symmetry
can be described by universal scaling functions with the help of two
characteristic  functions.
The one that arises from the 
off-critical temperature is given through the transcendent 
 equation (\ref{ee51}).  
The other one concerning the scaling property of $m_{0}$
is exactly the same as in the critical dynamics. This implies that
the characteristic function $\varphi$ does not depend on the
dynamics in the model under consideration.
If the non-linear part of a dynamics is skipped, one will always obtain 
a monotonously decreasing effective dimension for the initial order.
But it 
should not be a general conclusion.
For the 3-state Potts model, instead a monotonously decreasing function 
the effective dimension $x_{0}(m_{0})$ firstly 
increases then goes to zero \cite{s9}. 
Analytical calculations of 
characteristic functions from the Ising model
or the Potts model remain as a challenge.  

Although there is a similarity in the short-time behaviors between the 
SCD and POP, the physical pictures should
be quite different for the two processes. In the critical dynamics, the 
characteristic length is the correlation length which is divergent as
$t\rightarrow \infty$. Instead, it is the domain-size that 
tends to diverge in
the POP. The domain-width will be controlled by the
temperature. When the temperature approaches $T_{c}$ from below, the 
domain-width grows as $t_{\mu}^{1/z}$. As it 
 gradually become comparable with the domain-size, the 
phase-ordering picture ceases to be valid. 
From the consistency condition of the solution for $m(t)$, 
we obtain the boundary temperature $T_{b}$ for the ordering scaling domain.

In the specific example of $k=3/2$, $m(t)$ is obtained for all temperatures. 
The ordering scaling can smoothly cross over 
to the critical scaling and then to the high-temperature phase. But at the
crossover temperatures, there is no simple scaling law. 

We find that the bulk time-dependent order has the special feature that it
depends on $d$ and the interaction range-parameter $\sigma$ only through
their ratio. The longer (shorter) the
interaction-range is, the higher (lower) the effective dimension. It is physically
plausible because the correlation can be strengthened (weakened) 
by either increasing (decreasing) the interaction-range or dimension.

The LLNM should have the same  properties
as those we find in the KSM.
Therefore our results about the 
characteristic functions and the intermediate-time scaling in the POP
can be used as
the starting-point of finite-$n$ expansions.

We have investigated the KSM in the non-classical regime $d/2<\sigma<d$. 
In the classical regime $0<\sigma<d/2$, the initial increase would
be un-observable. When $\sigma\geq d$, the situation would be more complicated.
At $\sigma=d$, there would be a logorithmic correction. We leave these for further
investigation. Our results can also be extended to a finite-size lattice
\cite{s15, s15b}.

\vskip 1.pc
{\bf Acknowledgements:}Authors thank the very helpful discussions with B. Zheng
and L. Sch\"ulke.

\vskip 2.pc

\begin{figure}[p]\centering
\epsfysize=12cm
\epsfclipoff
\fboxsep=0pt
\setlength{\unitlength}{1cm}
\begin{picture}(13.6,12)(0,0)
\put(0,0){{\epsffile{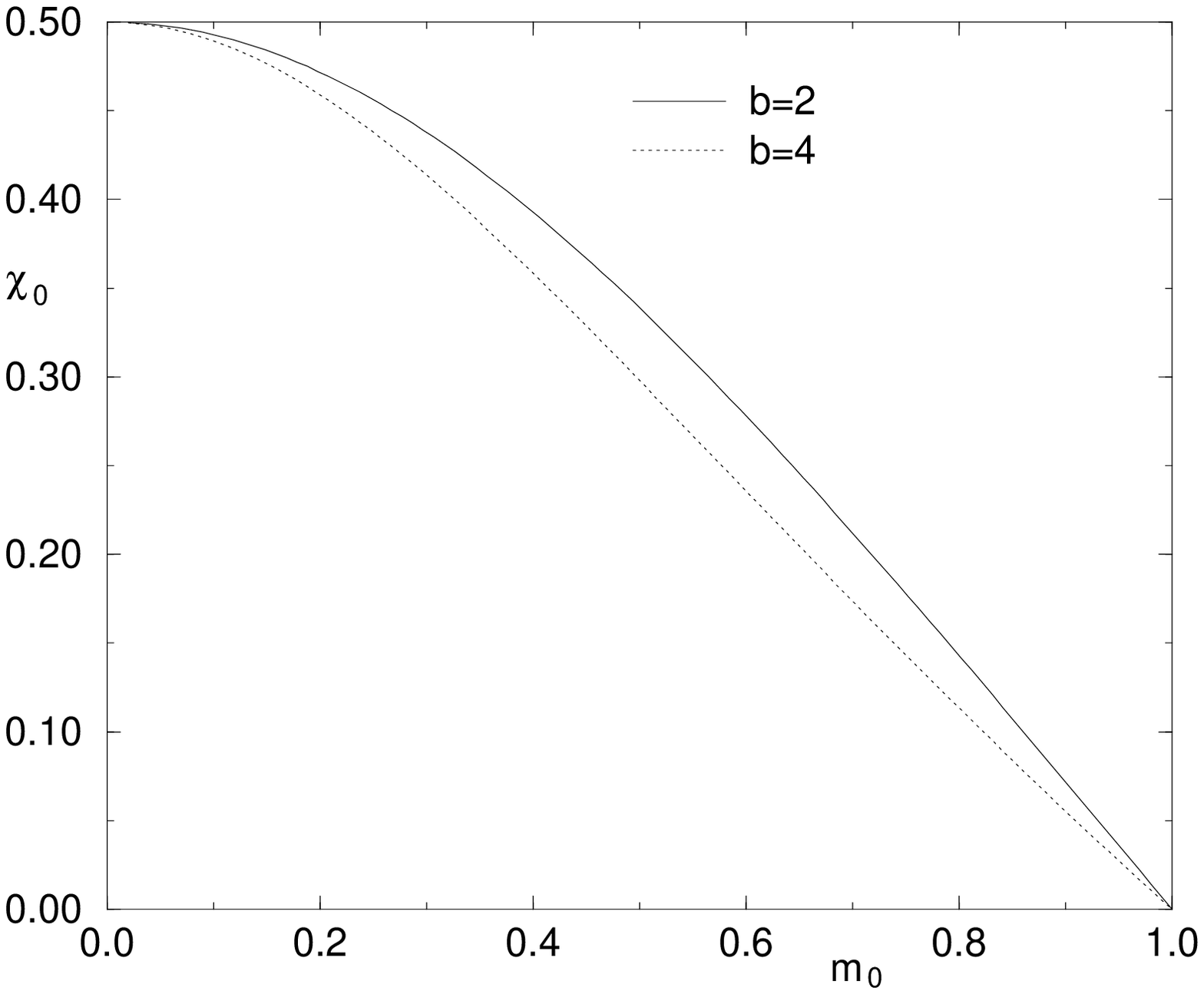}}}
\end{picture}
\caption{The effecitive dimension $x_{0}(b,m_{0})$  versus $m_{0}$.
The upper line is for $b=2$ while the lower line for $b=4$.
}
\label{f1}
\end{figure}

\begin{figure}[p]\centering
\epsfysize=9cm
\epsfclipoff
\fboxsep=0pt
\setlength{\unitlength}{1cm}
\begin{picture}(13.6,12)(0,0)
\put(0,0){{\epsffile{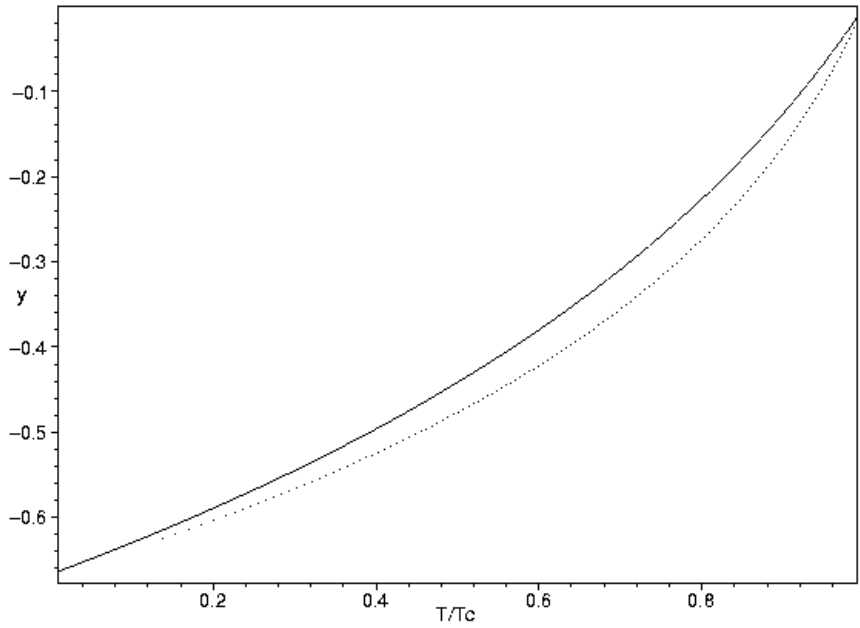}}}
\end{picture}
\caption{The effecitive dimension $y(b,T/T_{c})$  versus $T/T_{c}$ with
$k=3/2$.
The upper line is for $b=2$ while the lower line for  $b=4$.
}
\label{f2}
\end{figure}

\begin{figure}[p]\centering
\epsfysize=9cm
\epsfclipoff
\fboxsep=0pt
\setlength{\unitlength}{1cm}
\begin{picture}(13.6,12)(0,0)
\put(0,0){{\epsffile{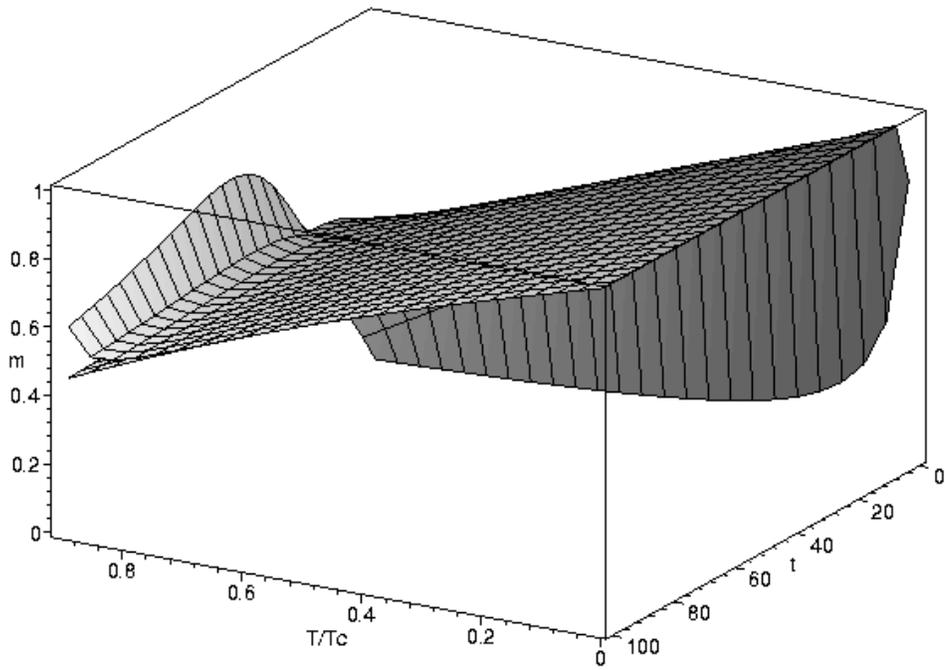}}}
\end{picture}
\caption{ The time-dependent order $m(t)$ with $k=3/2$ of (39) and 
(53). In this figure, $t_{i}$ has been used as the time scale.
The surface has infinite singularity in larger temperatures is the results of
(39) obtained by the low-temperature expansion, the other is the exact
 results of (53).
}
\label{f3}
\end{figure}

\end{document}